\documentclass[a4page,12pt]{article}
\usepackage{graphicx, epsfig, fancyhdr, layout, vmargin, exscale}
\usepackage[small]{caption2}
\usepackage{subfigure}
\usepackage{latexsym}
\begin{document}

\begin{center}

{\large \bf On spontaneous symmetry breaking in hot
QCD}\footnote{Talk at the conference "Fizika-2005" dedicated to
the 60th anniversary of National Academy of Sciences of
Azerbaijan;
Institute of Physics, Baku, 7-9 June 2005.}\\
\vspace{5 mm} {\bf Fuad M. Saradzhev}\footnote{e-mail:
fuad\_saradzhev@hotmail.com}\\
\vspace{5 mm}
{\it Institute of Physics, National Academy of Sciences of Azerbaijan,\\
H.Javid pr. 33, AZ-1143 Baku, Azerbaijan}\\
\vspace{5 mm} \bf{Abstract}
\end{center}
We prove that nontrivial vacuum states which can arise in hot QCD
are associated with the tachyonic regime of hadronic matter
fluctuations. This allows us to improve the condition for such
states to appear.

\vspace{1 cm} 1. It is known that at phase transitions from
hadronic to quark and gluon degrees of freedom nontrivial local
vacuum states can appear in the hadronic phase \cite{pis98}. These
states are metastable and of particular interest since they have
experimental signatures such as an enhanced production of $\eta$
and ${\eta}^{\prime}$ mesons \cite{kap96}. They can decay via CP
violating processes such as ${\eta} \to {\pi}^0 {\pi}^0$ and
because of global parity odd asymmetries for charged pions. The
decay rate of CP-odd metastable states was estimated in
\cite{ahren00}.

In \cite{blas02} we used the mean-field approximation to develop
the kinetic approach to the decay of the CP-odd phase in hot QCD
and to derive a non-Markovian kinetic equation describing the
production of ${\eta}^{\prime}$-mesons. A different kinetic
equation was derived for the production of tachyonic modes
\cite{sar03}.

In the present Talk, we aim to show that in addition to these
metastable states nontrivial vacua can appear according to the
standard spontaneous symmetry breaking picture provided the
hadronic matter fluctuations enter a tachyonic regime.

2. We start from the singlet Witten-DiVecchia-Veneziano effective
Lagrangian density \cite{vene79}
\begin{equation}
{\cal L} = \frac{1}{2} \Big( {\partial}_{\mu} {\eta} \Big) \Big(
{\partial}^{\mu} {\eta} \Big) + f^2 {\mu}^2 \cos\Big(
\frac{\eta}{f} \Big) - \frac{a_0}{2} {\eta}^2, \label{1}
\end{equation}
where $f=\sqrt{\frac{3}{2}}f_{\pi}$ and $f_{\pi}=92 MeV$ is the
semileptonic pion decay constant; ${\mu}^2=\frac{1}{3}(m_{\pi}^2 +
2m_{K}^2)$ is a parameter depending on ${\pi}$- and $K$-meson
masses. The parameter $a_0$ represents the topological
susceptibility. For zero temperature $T=0$,
$a_0=m_{\eta}^2+m_{{\eta}^{\prime}}^2 -2m_{K}^2 \simeq 0.726
{GeV}^2$, ${\mu}^2 \simeq 0.171 {GeV}^2$ and $f_{\pi} \simeq 93
MeV$. In response to non-zero temperature mesons change their
effective masses, ${\mu}$ and $a_0$ becoming functions of $T$. The
model is defined in a finite volume: $-L/2 \leq x_i \leq L/2$,
$i=1,2,3$. The continuum limit is $\frac{1}{V} \sum_{\vec{k}}
\Rightarrow \int \frac{d^3 \vec{k}}{(2{\pi})^3}$.

The meson field ${\eta}(\vec{x},t)$ obeys the Klein-Gordon type
equation
\begin{equation}
\Big( \Box + m_0^2 \Big) {\eta} =J_s, \label{2}
\end{equation}
where $m_0^2 \equiv a_0 + {\mu}^2$ and the current
\begin{equation}
J_s \equiv -f{\mu}^2 \Big[ \sin\Big( \frac{\eta}{f} \Big) - \Big(
\frac{\eta}{f} \Big) \Big] \label{3}
\end{equation}
is non-linear in $\eta$, i.e. contains orders ${\eta}^3$ and
higher and is therefore completely determined by the
self-interaction of the field $\eta$.

Following the mean-field approximation we decompose
${\eta}(\vec{x},t)$ into its space-homogeneous vacuum mean value
${\phi}(t)=\langle {\eta}(\vec{x},t) \rangle$ and fluctuations
$\chi$
\begin{equation}
{\eta}(\vec{x},t) = {\phi}(t) + {\chi}(\vec{x},t), \label{4}
\end{equation}
with $\langle {\chi}(\vec{x},t) \rangle=0$. The vacuum mean field
is treated as a classical, self-interacting background field. It
is defined with respect to the in-vacuum $|0 \rangle$ as
\begin{equation}
{\phi}(t) \equiv \langle {\eta}(\vec{x},t) \rangle \equiv
\frac{1}{L^3} \int d^3x \langle 0|{\eta}(\vec{x},t)|0 \rangle,
\label{5}
\end{equation}
so in the limit $t \to -\infty$ ${\phi}(t) \to 0$, while quantum
fluctuations take place at all times.

Substituting Eq.(4) into Eq.(3) yields the following decomposition
for the current
\begin{equation}
J_s=J_s^{(1)} + \overline{J}_s, \label{6}
\end{equation}
where
\begin{equation}
J_s^{(1)} \equiv J_s^{(0)} + {\mu}^2 \Big[ 1 - \cos\Big(
\frac{\phi}{f} \Big) \Big] {\chi} \label{7}
\end{equation}
is the current in the first order in $\chi$ with the background
field-fluctuations interaction term added, while the zero order of
the current
\begin{equation}
J_s^{(0)} \equiv -f{\mu}^2 \Big[ \sin\Big(\frac{\phi}{f}\Big) -
\Big(\frac{\phi}{f}\Big) \Big] \label{8}
\end{equation}
represents only the self-interaction of the background field. The
second current in the right-hand side of Eq.(\ref{6}) includes
terms of second and higher orders in $\chi$
\begin{equation}
\overline{J}_s=-f{\mu}^2 \sin\Big(\frac{\phi}{f}\Big)
\Big[\cos\Big(\frac{\chi}{f}\Big)-1\Big] -f{\mu}^2
\cos\Big(\frac{\phi}{f}\Big)\Big[\sin\Big(\frac{\chi}{f}\Big)
-\Big(\frac{\chi}{f}\Big) \Big]. \label{9}
\end{equation}
Substituting Eq.(\ref{4}) also into Eq.(\ref{2}) and taking  the
 mean  value $\langle ... \rangle$ yields the vacuum mean field
equation
\begin{equation}
\ddot{\phi}+a_0{\phi}+f{\mu}^2 \sin\Big(\frac{\phi}{f}\Big)
=\langle \overline{J}_s \rangle. \label{10}
\end{equation}
Eq.(\ref{10}) is a generalization of the vacuum mean field
equation used in \cite{sar03} for non-vanishing values of $\langle
\overline{J}_s \rangle$ (see also \cite{sar05}). In the
Hartree-type approximation,
\begin{equation}
\langle \overline{J}_s \rangle=-f{\mu}^2
\sin\Big(\frac{\phi}{f}\Big) \langle
\cos\Big(\frac{\chi}{f}\Big)-1 \rangle. \label{11}
\end{equation}
The equation of motion for the quantum fluctuations reads
\begin{equation}
\Big( \Box + m_{eff}^2 \Big){\chi}=\overline{J}_s - \langle
\overline{J}_s \rangle \label{12}
\end{equation}
with
\begin{equation}
m_{eff}^2 \equiv a_0 +{\mu}^2 \cos\Big(\frac{\phi}{f}\Big).
\label{13}
\end{equation}
For $a\equiv ({a_0}/{{\mu}^2})<1$, $m_{eff}^2$ can be negative for
some values of the background field indicating a tachyonic regime.
Eqs.(\ref{10}) and (\ref{12}) are self-consistently coupled and
include back-reactions. The vacuum mean field modifies the
equation for fluctuations via a time-dependent frequency, while
the fluctuations themselves react back on the vacuum mean field
via the source term $\langle \overline{J}_s \rangle$.

3. With the decomposition (\ref{4}), we deduce from (\ref{1}) the
effective Lagrangian density governing the dynamics of
fluctuations
\[
{\cal L}_{\chi} = \frac{1}{2} \Big( {\partial}_{\mu} {\chi} \Big)
\Big( {\partial}^{\mu} {\chi} \Big) + f^2 {\mu}^2 \cos\Big(
\frac{\phi}{f} \Big) \cdot \Big[ \cos\Big( \frac{\chi}{f} \Big) -
1 \Big]
\]
\begin{equation}
- f^2 {\mu}^2 \sin \Big( \frac{\phi}{f} \Big) \cdot \Big[
\sin\Big( \frac{\chi}{f} \Big) - \Big( \frac{\chi}{f} \Big) \Big]
- \frac{a_0}{2} {\chi}^2 - \langle \overline{J}_s \rangle {\chi},
\label{14}.
\end{equation}
Expanding (\ref{14}) in power series in ${\chi}$, yields in the
second order
\begin{equation}
{\cal L}_{\chi}^{(2)} = \frac{1}{2} \Big( {\partial}_{\mu} {\chi}
\Big) \Big( {\partial}^{\mu} {\chi} \Big) - \frac{1}{2} m_{eff}^2
{\chi}^2. \label{15}
\end{equation}
For $a>1$, the second order effective potential of fluctuations
\begin{equation}
V_{\chi}^{(2)} = \frac{1}{2} m_{eff}^2 {\chi}^2 \label{16}
\end{equation}
is ${\chi}^2$-type potential with oscillating walls. During the
time evolution of the background field, the potential (\ref{16})
fluctuates around $\frac{1}{2} a_0 {\chi}^2$ in tune with the time
dependence of ${\phi}$. For $a<1$, for some values of the
background field the potential (\ref{16}) becomes upside down
without any stable, particle states.

Let us consider now the exact form of the effective potential,
\[
\overline{V}_{\chi} \equiv \frac{1}{f^2{\mu}^2} V_{\chi} =
\frac{a}{2} \Big( \frac{\chi}{f} \Big)^2 + \frac{1}{f{\mu}^2}
\langle \overline{J}_s \rangle \Big(\frac{\chi}{f}\Big)
\]
\begin{equation}
- \cos\Big( \frac{\phi}{f} \Big) \cdot \Big[ \cos\Big(
\frac{\chi}{f} \Big) - 1 \Big] + \sin\Big( \frac{\phi}{f} \Big)
\cdot \Big[ \sin\Big( \frac{\chi}{f} \Big) - \Big( \frac{\chi}{f}
\Big) \Big]. \label{17}
\end{equation}
It also changes during the time evolution of ${\phi}$. First of
all, the term $\frac{1}{f{\mu}^2} \langle \overline{J}_s \rangle
\Big( \frac{\chi}{f} \Big)$ shifts the minimum of
${\chi}^2$-potential from ${\chi}=0$ to ${\chi}= - \frac{\langle
\overline{J}_s \rangle}{a_0}$,
\begin{equation}
\frac{a}{2} \Big( \frac{\chi}{f} \Big)^2 + \frac{1}{f{\mu}^2}
\langle \overline{J}_s \rangle \Big(\frac{\chi}{f}\Big) =
\frac{a}{2f^2} \Big( {\chi} + \frac{\langle \overline{J}_s
\rangle}{a_0} \Big)^2 + . . . , \label{18}
\end{equation}
the coordinate of the minimum oscillating in tune with the
background field.

In addition, in the tachyonic regime the effective potential
exhibits the spontaneous symmetry breaking. Let us compare the
form of (\ref{17}) for two different values of the background
field, ${\phi}=2{\pi}$ and ${\phi}={\pi}$. For both values,
$\langle \overline{J}_s \rangle =0$ in the Hartree-type
approximation. For ${\phi}=2{\pi}$, the effective potential takes
the form
\begin{equation}
\overline{V}_{\chi} = \frac{a}{2} \Big( \frac{\chi}{f} \Big)^2
-\cos\Big( \frac{\chi}{f} \Big) + 1. \label{19}
\end{equation}
It is positive for all values of ${\chi}$ and its minimum is at
${\chi}=0$.

For ${\phi}={\pi}$,  the effective potential becomes
\begin{equation}
\overline{V}_{\chi} = \frac{a}{2} \Big( \frac{\chi}{f} \Big)^2 +
\cos\Big( \frac{\chi}{f} \Big) - 1. \label{20}
\end{equation}
It is minimized for
\begin{equation}
a \Big( \frac{\chi}{f} \Big) = \sin\Big( \frac{\chi}{f} \Big).
\label{21}
\end{equation}
For $a \geq 1$, Eq.(\ref{21}) has only trivial solution
${\chi}=0$. However, for $a <1$ nontrivial solutions appear.

\begin{figure}[tbp]\centering
\includegraphics[clip, angle=-90, width=12.3cm]{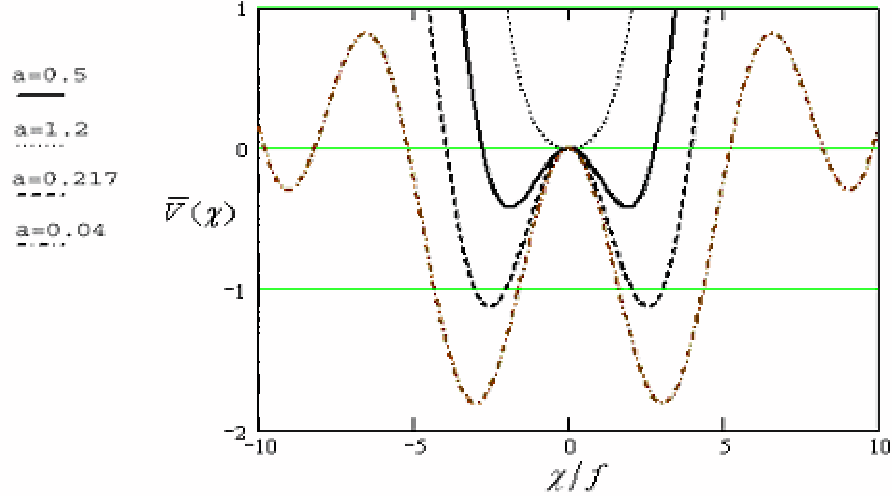}
\caption{\it The shape of the potential $\overline{V}(\chi)$ for
different values of $a$.} \rm  \label{fig:1} \rm
\end{figure}

Fig.(\ref{fig:1}) shows the shape of the potential
$\overline{V}_{\chi}$ for different values of $a$. The nontrivial
local minima appear for $a<1$. For $a>1$, the spontaneous symmetry
breaking does not occur. The special value $a_{sp}=0.217$ was
found in \cite{pis98}. For $a<a_{sp}$, the number of nontrivial
local minima is increasing with decreasing values of $a$.The
nontrivial minima are of different energy; the ones of higher
energy are metastable and can decay by a tunneling.

4. The tachyonic regime can be characterized as a regime of
spontaneous symmetry breaking. Although $a_{sp}=0.217$ is
specified in \cite{pis98} as a special value defining the first
local minima, we have shown that nontrivial minima appear even for
$a_{sp}<a<1$.

Whether the system evolves in the tachyonic or non-tachyonic
regime is fixed by the value of the background field. During its
time evolution, energy is transferred from ${\phi}$ to $\chi$. As
a result, $\phi$ is damped, while the number of particles in
quantum fluctuations increases. If, for example,
${\phi}(t=0)=2{\pi}$, then the quantum fluctuations first evolve
in the standard, non-tachyonic regime. As soon as ${\phi}(t)$
reaches $\pi$, the tachyonic regime starts (for $a<1$), and the
intensive production of tachyonic modes results in a rapid damping
of $\phi$.

\end{document}